# Enhancing students' learning process through self-generated tests

Marcos Sanchez-Elez, Inmaculada Pardines, Pablo Garcia, Guadalupe Miñana, Sara Roman, Margarita Sanchez, Jose Luis Risco

*Universidad Complutense de Madrid (SPAIN)*

marcos@fis.ucm.es

**Abstract.** The use of new technologies in higher education has surprisingly emphasized students' tendency to adopt a passive behavior in class. Participation and interaction of students are essential to improve academic results. This paper describes an educational experiment aimed at the promotion of students' autonomous learning by requiring them to generate test type questions related to the contents of the course. The main idea is to make the student feel part of the evaluation process by including students' questions in the evaluation exams. A set of applications running on our university online learning environment has been developed in order to provide both students and teachers with the necessary tools for a good interaction between them. Questions uploaded by students are visible to every enrolled student as well as to each involved teacher. In this way, we enhance critical analysis skills, by solving and finding possible mistakes in the questions sent by their fellows. The experiment was applied over 769 students from 12 different courses. Results show that the students who have actively participated in the experiment have obtained better academic performance.

## Introduction

The interactions of technology and education may be centrally important in shaping education in the future (Monfort & Brown2012). Teaching/learning strategies are being deeply influenced by the use of new technologies. Modern universities offer online learning environments as a complement to traditional education methods in order to help students in their learning process (Oncu & Cakir, 2011). This technology allows students to download class notes at home, discuss their ideas in forums, upload class assignments, check their knowledge by online tests, etc.

An increasing number of graduate students are nowadays using laptops, tablets or even smartphones as their basic equipment, which emphasizes the use of online learning environments. The number of students attending lessons with digital devices is so common that teachers have started to debate upon this having an adverse effect on their learning process or not. Several studies have been made on this issue (Fried, 2008; Carrick-Simpson & Armatas, 2003; Yarnall et al. 2006).

New technologies are also part of the daily lectures given in our teaching rooms: audio, video, PowerPoint presentations, real-time web interaction, etc. Despite how useful and innovative these methods are, we have noted that they turn the student into a passive information receiver. All the information provided during lectures is easily accessible through the university online learning environment. Therefore they do not show much interest in taking notes, which makes knowledge a very volatile acquisition. This makes it really hard for the teacher to maintain continuity in the development of the course. Apple Inc. has noticed this students' passivity over the world and is promoting innovative educational initiatives to enhance active skills in students (Apple, 2011).

In this regard, we propose ways to use current technologies, helping students to participate actively in the learning process. Barak (2004) provides evidence that web-based activities can serve as both learning and assessment enhancers in higher education by promoting active learning, constructive criticism and knowledge sharing. In addition, the technology can be used to promote students intrinsic motivation. This type of motivation is defined by Ryan and Deci (2010) as the interest of doing an activity for the inherent satisfaction of the activity itself.

In addition, we fully agree that the act of 'composing questions' focuses the attention of students on content, main ideas, and checking if content is understood (King 1994, Rosenshine et al. 1996). This helps students to acquire a deeper level of knowledge instead of learning by heart, since the student who formulates a question needs to understand the subject (Marton & Säljö, 1984).

In this context, this paper presents a methodology to encourage students' capacity to compose questions by taking advantage of the Information Technologies (ITs). We have developed a software tool that automatically generates evaluation tests mixing the questions generated by the students with the teacher's questions. This tool has been integrated into the 'Universidad Complutense de Madrid' online learning environment (Moodle and Sakai platforms) so that any teacher in the university may use it for automatic test generation.

This provides the teacher with a tool to obtain a quantitative evaluation of the knowledge acquired by individual students together with a qualitative feedback of the impact of the lectures on the students. It is not our aim to compete with any existing tools, but only to provide the means for teachers to effectively improve students' motivation for active learning.

The use of the questions generated by the students to create the class test has two main effects: First, it promotes debate over these questions among them. Students are encouraged to discriminate between relevant and accessory information and as a consequence, good management of information skills are developed. Moreover, it is not threatening for the students to write questions even if they are not sure about the quality, since they are not graded for the quality of their questions. Second, they are intrinsically motivated to remain actively involved in the methodology, as they find that the questions they have proposed or they have discussed on-line appear in the test class. Third, regardless of whether one is teaching in a traditional or an active learning class, students' questions provide an opportunity to understand what students have in mind and help to uncover their misconceptions and/or preconceptions (Marbach-Ad 2000).

This methodology has been designed to be helpful in the process of acquiring the contents of the tough subjects. It encourages the development of desirable skills in all university graduates such as content synthesis and analysis, good management of information and proficiency in written language abilities.

In this paper we analyze the effectiveness of the proposed methodology in terms of academic performance by comparing that of the students who participated and that of those who did not. We are aiming at applying this method in a wider range of courses in the coming years. To date, the experiment was performed in different under-graduate courses in our university and has proved successful in encouraging student participation in the teaching-learning process.

This methodology arose in response to a common behavior that has been observed in Computer Science students, especially in the first and second courses. A high percentage of them presented a very passive attitude during the lessons, did not participate in the academic activities and works proposed by teachers, leaving all his learning effort for the final examinations.

With the implementation of the Bologna system of education (Keeling, 2006), a new type of student is expected. The new degrees revise the process of learning and programs contents. Then, an active student involved in the academic activities is needed. In fact, the Bologna directives have imposed a new evaluation method where the final exam is not enough to pass a subject, and a continuous work is required. This evaluated as part of the final qualification, in conjunction with the final exam.

The results obtained, in our School, by the five groups of the first year in this new framework, are not as good as the teachers' effort and illusion deserve. A low percentage of pupils, range from 50% to 60%, was able to obtain the minimum qualification to pass a subject, this is due in part to a very low presence of students in classroom. These results lead directly to the question of how teachers may motivate their students, if this is possible, to participate in the academic activities and to follow the evaluation process until the end of the academic year.

## Related Work

Traditional methodology to help student to acquire new information and connect new concepts consists on teachers' questions and student discussions. The teacher usually asks questions during lectures and/or creates tests to be solved by the students during lecture time. Questions allow a teacher to determine how well the topics have been learnt and whether there is need for additional instruction. However, educators must invite students to experience the world's richness, empower them to ask their own questions and seek their own answers, and challenge them to understand the world's complexities (Brooks et al. 1993). There are several works trying to shift this responsibility to students (King 1994, Rosenshine et al. 1996). These authors analyze how questions generated by students about the topics they had read resulted in gains in comprehension. A number of studies are focused on promoting question generation associated to text comprehension (Cohen 1983; Dreher and Gambrel 1985; Van den Broek et al. 2001). A more elaborate study about the same hypothesis, but centered on science education, can be found in Chin and Brown (2002); Taboada and Guthrie (2006) and Costa (2000). Raeff (2009) believes that both individual and social progress is to a great extent based on the ability to ask questions and answer them rationally. The methodology described in the papers above mentioned share the same basic principles of our proposal. However, our work is focused on encouraging the participation of students in the process of building questions by taking advantage of the use of information technologies, and not on the quality of the questions or on guiding the questions generation process.

The use of web-based learning environments improves students learning, simplifies the assessment process and provides teachers with feedback to emphasize the misunderstood concepts (Yu-Feng et al. 2012). The methodology proposed uses the tools available in any online learning environment, such as chats and web-forums, to promote collaborative learning (Harasim 1996). Collaborative learning has generally shown that student interaction may increase group performance and individual learning outcomes.

There are several approaches that promotes online collaborative learning (Rao and Collins 2002; Slusser and Erickson 2006). These proposals try to enhance active learning through tests the students solved in groups. However, authors in (Leight et al. 2012) affirm that this approach improves performance, but not content retention. Our approach uses a different perspective; we propose a collaborative strategy where the teacher makes up an online test with the questions created by the students. This test will be available for the class for a certain time. They can use it as a learning tool, to study the subject alone or in group, but the classroom tests are solved individually.

We also may find methodologies to increase active learning by asking students to create questions and to upload them to a web-based environment (Pollard 2006; Fong-Ling et al. 2009). Both papers describe a methodology based on collaboration (students work in groups) and competition between groups (the winner group receives a prize). The objective of our proposal is to create a collaborative learning environment that helps students to consolidate the knowledge acquired during lessons instead of encouraging competition.

Another approach to an online student question-generation learning system is proposed in Fu-Yun (2011). In this work students create questions which are assessed by their classmates. The quality of the questions is evaluated, and the assessors can include comments to the authors. The system is provided with the possibility that the authors can defend their reasoning, establishing a constructive dialogue between both parts. This interaction has an influence on the knowledge building process and on the final outcomes. In our proposal, questions created by the students are not evaluated but corrected in forums by classmates if they are incorrect or mis-formulated. Our study demonstrates that regardless of the quality of generated questions and the degree of participation, all the students (even the passive ones) improve their academic results.

When the students have to work through the subject in an online environment wherein they have to formulate their own questions, they may have difficulty in changing dependent learning habits; those problems can arise if students are not self-motivated (Taplin 2000). As a result, the student may give up in getting involved in this type of learning methodology. When a strategy is formulated, it has to take into account the students' intrinsic motivation. There are different proposals of e-learning environments which aim to increase the intrinsic motivation. Most of them are designed as a game-like realistic simulation in which students play a certain role to acquire the concepts (Martens et al. 2004) or to enhance their creativity (Leng et al. 2010). However, our proposal tries to exploit the tendency of young people to use social networks. Students can study a subject by themselves and afterwards work and discuss the subject with their classmates in the forums of the e-learning environment.

There are previous works focused on this context that try to improve active learning using social networks like Twitter or Facebook. These studies are based on the idea that to improve the learning performance it is necessary to increase the discussion and participation of students during lectures. In Tiernan (2013), the author proposed the use of Twitter to fully engage students in lectures. The results of this approach are positive since it increases the participation of students compared to the traditional methodology. Shy students or those with lack of self-confidence have a new opportunity to interact with their fellows. In another work, Facebook is used to increase the engagement of first-year law undergraduates in Australian University (Coldwell et al. 2011). Through Facebook students dare to participate, formulate and answer questions about lecture topics. The teacher can get feedback on students' understanding of the lectures and which concepts need to be reviewed.

## Methodology

Our experience as teachers during the last years has moved us to search for learning methods that could motivate students to participate actively in the development of their learning process. In this section the applied methodology is described. As mentioned above, our method is focused on the development of autonomous learning strategies for students. To reach this goal, we have developed a method based on continuous assessment by performing several tests throughout the course, where tests are composed by questions created by both students and teacher.

The methodology consists of three main stages (summarized in Table 1), which are described below:

Questions collecting: once the teacher has explained a module of the course and has performed several exercises, a period of one-week is opened for the students to upload multiple-choice test questions related to the module. We have observed that the process of creating test questions is initially a hard task for our students. Thus, the teacher facilitates an initial set of several solved questions. This initial set must serve as a guide, i.e., the teacher leads the students to be creative by defining questions different to those provided in the initial example.

It is worth highlighting that the quality of the questions is not directly analyzed. This decision has been taken to encourage students to post questions. However, students can discuss about their quality during tasks II and III described in Table 1. In addition, students can also indicate if both the statement of the question or the answer is incorrect.

First draft: after all the questions have been uploaded, they are published online for a another one-week period. At this time, all the students can see the questions and solve them. In this case students must check if questions are ambiguous or misleading, if questions should be revised or eliminated or if incorrect alternative answers should be revised or replaced. To this end, a forum is enabled to discuss and fix possible errors. This stage is also used to talk about particular doubts and difficulties with other students or with the teacher if he/she is available. It should be noted that, in this phase, the evaluation of the students' results is not necessary. We simply want students to practice and to learn using the questions uploaded by their classmates. In this way, students perceive that they participate in their grading process seeing that their questions may be part of a classroom test. At the same time, since students can see all the proposed questions, we stimulate critical analysis (the student must understand the quality of other questions), and adequate information management, mainly because in function of the quantity of proposed questions a student may need to use some selective strategy that allows him/her a better use of the available information.

Test examination: when the time to discuss and correct uploaded questions is finished, the teacher can generate the test to be performed by the students in class. To compose the test, a subset of the questions uploaded is used. The teacher must perform a final verification regarding correctness and convenience, trying to reach a good balance between quality, variety and final difficulty. To this end, the teacher may remove some questions as well as to create some extra ones if he/she believes that uploaded questions by students do not cover the current teaching module.

Finally, there is one additional stage, which is not specific to the methodology: students have to pass a final regular exam covering the complete course syllabus (Task V). Generally, this exam is not a multiple choice test, but it has problems and application questions. We wish to emphasize the fact that making online and classroom tests during the course is a way to get students motivated to learn. Therefore, the feedback effect of the methodology helps assimilate basic concepts.

Following this method, a student may perform up to four tasks, as shown in Table 1.

**Table 1.** Summary of tasks that can be performed by students.

| Task | Description | Methodology Stage |
|---|---|---|
| Task I | Students send their questions and corresponding solutions | Questions collecting |
| Task II | Students talk about particular doubts, difficulties, etc. | Questions collecting |
| Task III | Students solve the initial evaluation test generated with their questions | First draft |
| Task IV | Final test using questions uploaded by the students and some extra questions defined by the teacher | Test examination |
| Task V | Students face the final exam, | Final exam |

Regarding to the impact of these tests on the course grades and the division of the subject into modules, our methodology establishes that a teacher may perform up to 5 tests per semester. Among these tests, the teacher selects the best 3 grades for each student, which may represent about a 15% of the final grade reached in the course by a student.

It is worth noting that even if a student is not involved in Task I, he/she can take advantage of performing Tasks II and III. These two tasks also motivate students to perform tests (Task IV) and to participate more actively in the course.

Following this methodology, we have observed several improvements with respect to the application of traditional tests, such as:

1) Students formulate questions about the content of the course. This point helps in a correct understanding of the main concepts.

2) They become more involved in course contents.

3) Communication between teachers and students is more fluid and regular and is not concentrated on the days before the final exam.

4) Collaborative learning between students is implicitly motivated (Boekaerts & Minnaert, 2003).

## Programs and tools associated with stages

To facilitate the tasks of our methodology, we have developed new software and we have used some existing tools as well. Next, we describe the integrated framework.

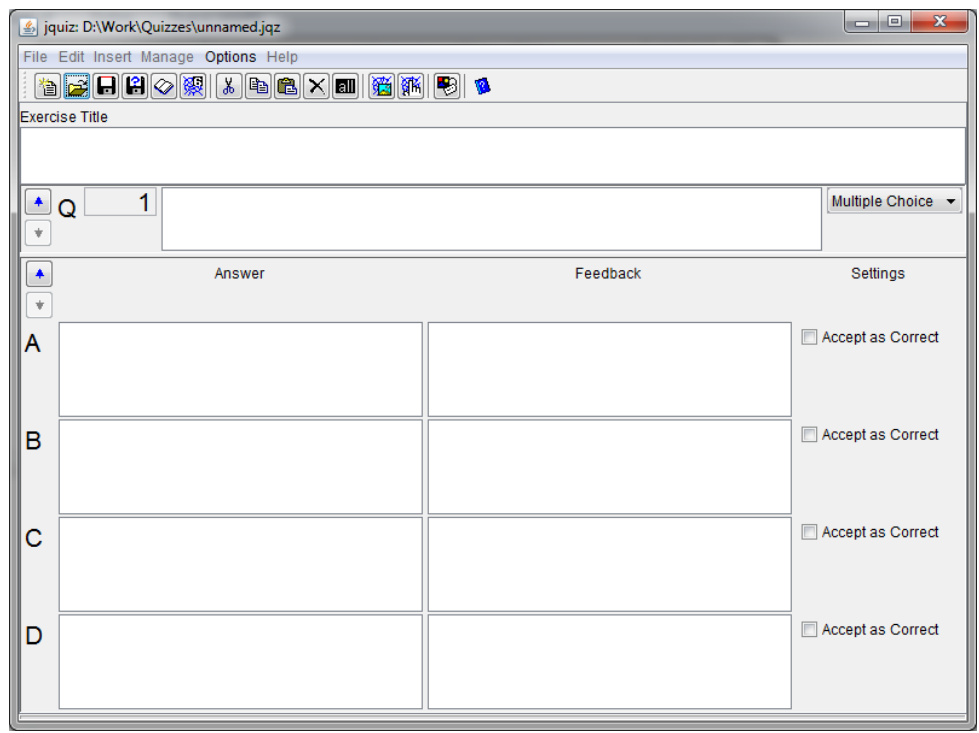

**Fig. 1**. Hot Potatoes graphical user interface to insert questions

## Questions collecting

At this point, every student may send his questions and corresponding solutions; currently, only multiple choice questions are allowed. We rely on the tool Hot Potatoes (HotPotatoes, 2012) to handle the test creation process. First, because it provides a graphical user interface, which prevents the students from committing formatting errors (see Figure 1) and, second, because it is freeware, and most of the e-learning environments include applications to transform the Hot Potatoes' format (jqz) to their own format (and vice versa).

As a result, every student who wants to contribute with one or more questions must send a jqz file (the output from Hot Potatoes, containing the tests) to the virtual campus.

## First draft and forums

The initial phase of this stage is performed by the teacher and consists on joining all the files received. To facilitate this job, we have developed a tool named HotPot2Pdf that is able to merge all the files received in a new jqz file. This tool has a "Concat jqz files …" button in the main window, the teacher clicks on it, and a browser appears where the teacher selects one of the jqz files within a directory. A new file with the name of the directory is created, that contains all the tests.

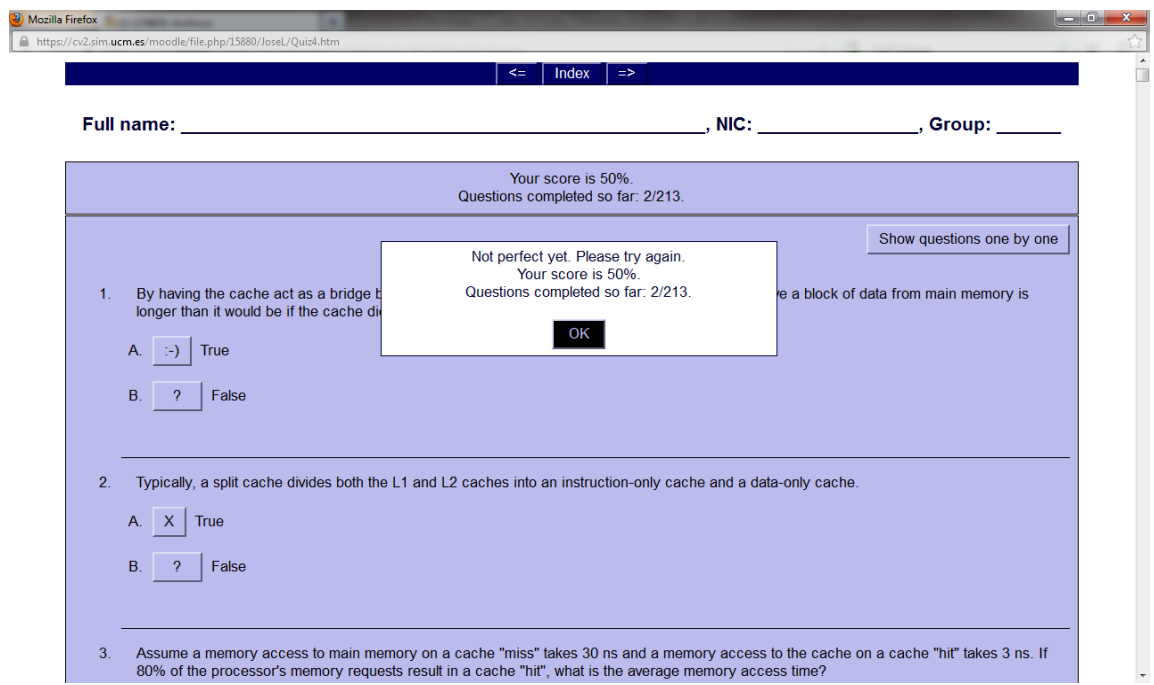

**Fig. 2**. Test in HTML with all the questions uploaded by the students. This test is made available online (via Moodle, for example).

There are mainly two ways to publish the test online: (1) to design an online test using the ability of the e-learning environment to import questions coming in Hot Potatoes' format and (2), to create an HTML test from Hot Potatoes, using the option "Make a Standard Exercise", which generates a HTML file that can be published via Moodle (Kuo et al. 2010). Figure 2 shows a set of final questions published in Moodle as HTML resource. As we can see in this figure, the total number of questions uploaded is 213, the student has correctly answered the first question and is failing in the second. In this particular case, the student can answer multiple times. However, this behavior can be configured in Hot Potatoes during the test generation phase.

**Test Examination**

The third stage in the proposed method is the examination. The students must complete a test elaborated by the teacher.

When performing the test in a classroom, one of the biggest problems encountered was the great amount of time needed to select random questions and to adequately format the text. Our tool, HotPot2Pdf, generates a test file in Portable Document Format (PDF), given an initial set of questions like.

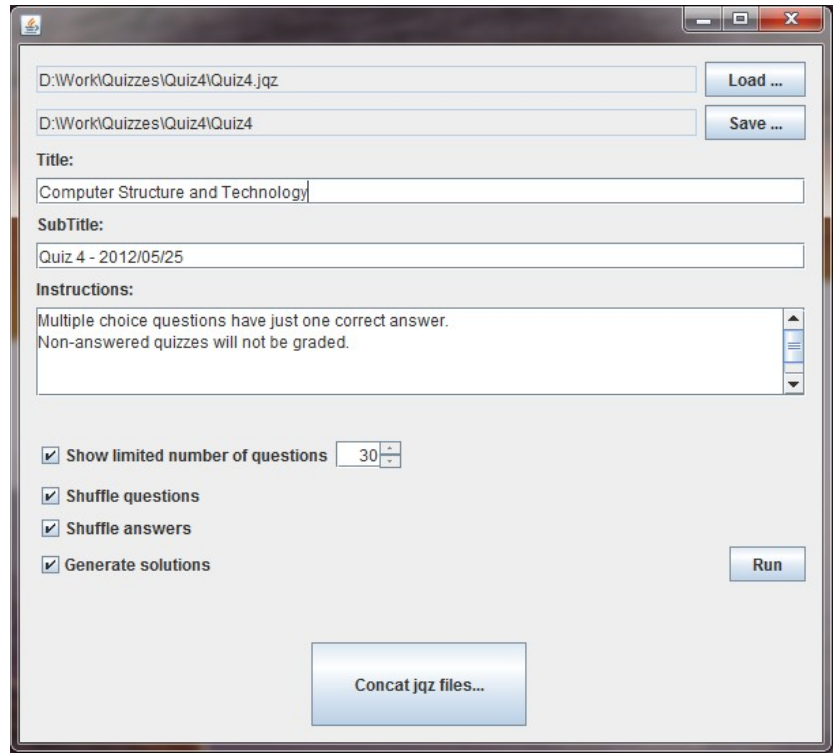

**Fig. 3**. Generating a test in PDF.

Figure 3 depicts the main window of the program HotPot2Pdf. Firstly, we must load the source file containing the set of questions from which we want to generate the test (Quiz4.jqz in the Figure), and the name of the generated PDF file we want to save (Quiz4.pdf). Furthermore, as Figure 4 shows, we may specify a title, subtitle, and instructions for the exam. Finally, we have the choice of generating a subset of questions (30 in this case), instead of including all the questions in the jqz file. We can also shuffle questions or answers, and generate a table at the end of the document with all the correct answers.

# Evaluation

**Experiment Scenario**

The methodology described in this paper was conducted on subjects that deal with issues related to electronics and hardware. Students who participate in the study were enrolled in different degrees (Computer Science, Computer Engineering, Electrical Engineering and Physics) at University Complutense of Madrid (UCM). Currently, UCM is the largest university in Spain. It offers nearly 80 majors, 230 individual degrees, and 221 PhD programs.

The experiment was designed including two groups that follow the methodology completely or partially (according to the tasks performed by the students, see Table 1). At the beginning of the course students can decide whether to participate or not by posting questions which determines their presence in a particular methodology group, that we named MG-I and MG-P. Moreover, the experiment has two control groups to compare our methodology to both traditional methods, with and without traditional tests, named CG and TG respectively. These control groups have been formed with students from the same subjects and years that the ones who participate in the experiment, but their teachers have not followed the methodology. These four groups are described below:

*Group CG (**Control Group**)* – participants from this group just do one final exam in order to pass the subject. This group is formed by 222 students of different years, as shown in Table 2.

*Group TG (**Tests Group**)* – participants from this group performed tests following the standard McDaniel's method (McDaniel Fisher (1991)). The tests and the material to prepare the tests were completely generated by the teachers. It is formed by 210 students of different years (see column TG in Table 2).

*Group MG-P (**Methodology Group with students Partially involved in the study**)* – participants from this group partially followed the proposed methodology. These students did not post any question, but they were involved in the study participating in the activities associated to the methodology (Tasks II, III, IV and V in Table 1). As Table 2 shows, this group is formed by 112 students of different years.

*Group MG-I (**Methodology Group with students fully Involved in the study**)* – participants from this group complete all the activities proposed in the methodology: they do the activities of the group MG-P and they generate and upload questions, i.e., tasks I, II, III, IV and V in Table 1. This group is formed by 225 students of different years.

**Table 2.** Number of students who have participated, classified in different groups

| Students course | CG | TG | MG-P | MG-I |
|---|---|---|---|---|
| First year | 59 | 71 | 36 | 63 |
| Second year | 82 | 139 | 64 | 118 |
| Third and Fourth year | 81 | 0 | 12 | 44 |
| Total | 222 | 210 | 112 | 225 |

All students of both MG groups were encouraged by the teachers to post test questions, as a result of the motivation, more than two thirds of students actively participated in the strategy (MG-I), as it can be seen in Table 2. We believe that the strength of the research design lies in the use of two control groups (CG and TG) to compare our methodology to both traditional methods: with and without traditional tests.

**Results**

*Participation of students in task I*

Task I in the methodology consists of sending questions and their corresponding solutions using an on-line learning environment. This task is only performed by the students in MG-I.

The degree of participation of students in the process of posting questions has not been uniform as shown in Table 3. There is a significant number of students that post only one question, but the largest group is represented by those who have generated between 2 and 4 questions.

**Table 3.** Statistics on the number of questions posted by the students to create test

| #Questions uploaded per test | Percentage of students |
|---:|---:|
| One | 33.81% |
| [2-4] | 43.64% |
| [5-10) | 9.80% |
| 10 or more | 12.75% |
| Mean | 2.49 |
| Standard Deviation | 5.14 |

*On grade distribution*

In this subsection we describe the grades obtained by the students of the four groups previously mentioned. Table 4 shows the data obtained. For each group, there are two columns: the first one indicates the number of students obtaining a certain grade, and the second one contains the percentage of students. In the means (and standard deviations) computation of the obtained grades we used the following equivalence: E and F correspond to 0 and A+ to 4, as explained in Appendix A (ECTS grade equivalence).

**Table 4.** Distribution of marks in all the groups and its statistics

|  | CG | | TG | | MG-P | | MG-I | |
|---|---|---|---|---|---|---|---|---|
| A+ | 2 | 1% | 6 | 3% | 7 | 6% | 18 | 8% |
| A | 4 | 2% | 15 | 7% | 15 | 13% | 45 | 20% |
| B and C | 46 | 21% | 32 | 15% | 22 | 20% | 49 | 22% |
| D | 60 | 27% | 69 | 33% | 37 | 33% | 57 | 25% |
| E and F | 110 | 49% | 88 | 42% | 31 | 28% | 56 | 25% |
| Means | 0.77 | | 0.96 | | 1.37 | | 1.61 | |
| Standard Deviations | 0.90 | | 1.06 | | 1.20 | | 1.27 | |

Students that follow the methodology completely are in the MG-I group (last two columns in Table 4). In this category, 50% of the students have achieved a grade of C or better, with a 28% of them reaching grade A or better. Over 75% of students have passed the subject.

In MG-P group over 72% of the students have passed the subject. In this category, 39% of students have a grade of C or better, whereas 19% have a grade A or better.

With respect to the control groups, in the CG group a 24% of students have a grade of C or better, with 3% of students having A or better. Meanwhile, in the TG group 25% of students have a grade of C or better, and 10% of them have A or better.

**Table 5.** Tukey Method. Analysis of the differences between groups with a confidence interval of 95%

| Contrast | Difference | Standardized difference | Critical value | Pr > Diff | Significant |
|---|---|---|---|---|---|
| MG-I vs CG | 0.843 | 8.052 | 2.569 | < 0.0001 | **Yes** |
| MG-I vs TG | 0.647 | 6.092 | 2.569 | < 0.0001 | **Yes** |
| MG-I vs MG-P | 0.234 | 1.827 | 2.569 | 0.260 | No |
| MG-P vs CG | 0.609 | 4.749 | 2.569 | < 0.0001 | **Yes** |
| MG-P vs TG | 0.413 | 3.189 | 2.569 | 0.008 | **Yes** |
| TG vs CG | 0.196 | 1.841 | 2.569 | 0.254 | No |

A direct comparison of the results presented in Table 4 suggests that the performance of MG groups is better than that of control groups (TG and CG). To confirm this hypothesis an analysis of variance (using ANOVA) together with a Tukey Method are performed (Table 5). This method consists of a single-step multiple comparison procedure and statistical test, which is used to find means that are significantly different from each other. There are significant differences for MG-I and MG-P versus TG and CG. However, there are not significant differences between both MG groups. Moreover the significance of the difference between MG-P and TG is clearly lower than between MG-I and TG.

Figure 6 depicts the contribution to the percentage of the marks obtained by the students for each group. First, almost 50% of students with A or A+ marks belong to the MG-I group (see "A+" and "A" bars in Figure 4). Second, almost 80% of the students with these highest scores participate in some way in the methodology, i.e., they belong to MG-I or MG-P. On the contrary, a third of the students who fail belongs to CG and the other third belongs to TG and just a 18% belong to MG-I. Finally, we may also highlight that more than 50% of the students with C or better marks belong to MG-P or MG-I.

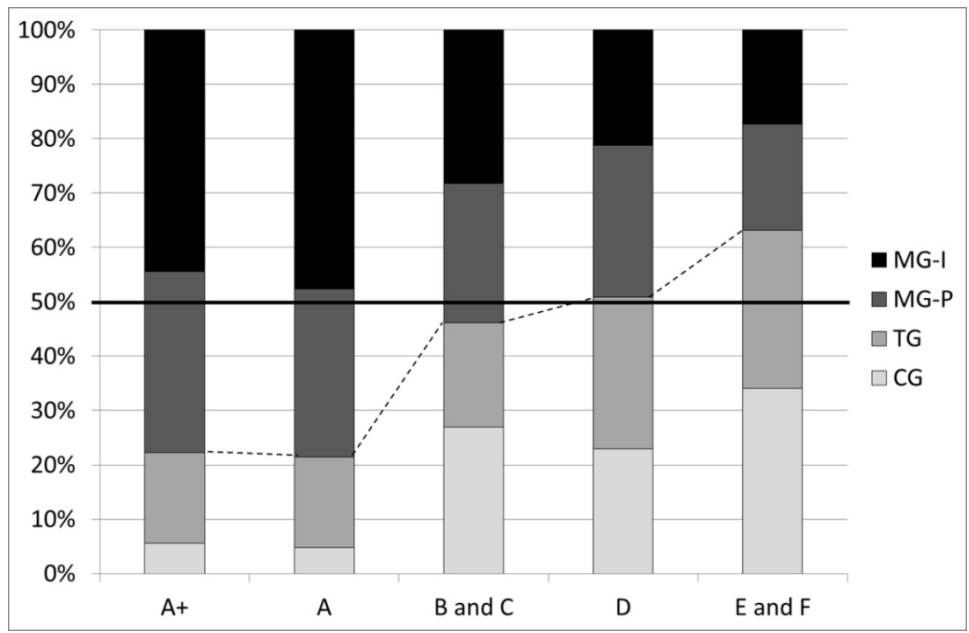

**Fig. 4**. Percentage of students per mark and group

## Results Discussion

In this section we analyze and compare the results obtained, to conclude that our methodology clearly improves the academic results of students.

A fair analysis of the strength of the proposed methodology is the number of students that complete the course (see Table 4). Comparing with CG, MG-I presents 24% more students that passed the course (21% more in MG-P). With respect to TG, MG-I presented 17% more students that passed the course (14% more in MG-P). Thus, our methodology improves the academic results of the students in MG groups, even if the student is not completely involved (as happens in MG-P). In both TG and MG-P the students have a passive role in the process of generating tests and, therefore, similar results might be expected. However, results are clearly better in the MG-P group.

Students often show unequal engagement in the collaborative learning activity; a few group members take the responsibility for the problem-solving, while others involve in social loafing and they are not motivated to interact with their partners (e.g. O'Donnell 1999). This is the main difference between the two MG groups, even though all students take advantage of the questions generated by their classmates. The collaboration among students is supported by the online publication of the questions and the discussions of their solutions in forums (Tasks II and III in Table 1), which helps MG-P students to pass the subject.

In order to measure the strength of the methodology we can also compare the percentage of students with the highest grades (A+ or A). The best results are achieved in MG-I group, where the learning methodology proposed in this paper is completely applied. The percentage of students with A+ or A in MG-I is 28%, versus 10% in TG or 3% in CG. It is remarkable that MG-P results (A+ and A are 19%) are also better than those obtained in the control groups, although not as good as those of MG-I. The group MG-P has, in percentage, double of students with the highest scores than TG, even though in both groups the work of the students is just test examination. From these excellent results, we infer two things: first, that all the students in MG-I and MG-P are motivated enough to work collaboratively getting the full benefit from the generated material (questions created by themselves, forums, online tests, …) and, second, that dedicating part of their study time to generate questions about the subjects helps the students to still achieve better marks (Rosenshine et al. 1996).

The qualitative description of the preceding paragraphs of the obtained results in the different groups explains the obtained results by applying ANOVA (Table 5). Both MG-P and MG-I have a significant difference from the control group (CG) in which students just face a final exam generated by the teacher (task V). The same applies to the comparison of the other control group (TG) versus the MG-P and MG-I groups. However, the differences between MG-P and TG are lower than between MG-I and TG. The reason is that students in these two groups do not participate in the process of generating questions. The difference between them lies in how they deal with the preparation of the subject, in the case of MG-P group this is more dynamic and collaborative.

For a better characterization of the MG-I group we present the number of questions posted by the students. These results are a measure of the motivation degree of the students who belong to this group. Then, from Table 3 follows that a clear majority of the students post less than five questions per test, and only 12.75% post ten or more questions. It is this subgroup of students which states the differences between MG-P and MG-I in obtaining high grades. However, the variances are not significant between the two groups, as indicated by the ANOVA analysis, because MG-I group also consists of students who do the least possible effort in this collaborative work, i. e., students that only post one question.

## Conclusions

During past years, we have noticed how the introduction of new technologies in the classrooms, is turning the students into mere information receivers, and it eventually influences negatively their academic performance. In order to modify this situation we have introduced a new educational methodology that uses the ITs to encourage students to play an active role in the learning process.

A software tool allows students to generate test type questions, related to the contents of the course, that will be part of the evaluation tests. Online forums supervised by the teacher serve to refine these questions and solve students' doubts. Besides, this framework promotes autonomous learning and increases the participation and interaction of students; essential to improve academic results.

In order to validate our new methodology, we applied it in some of our university courses. The initial results show that students who took part in the experiment obtained better grades. There are noticeable differences among those groups that follow standard methodologies and those groups that somehow take part in the experiment. The reason of this is that the students tend to be more critical, check the solution proposed, and works more on the subject, than when the test is created by the teacher. In addition, the methodology increases the sociability and interaction among classmates when studying the subject.

Furthermore, we can even observe a difference between those who participated actively, sending test questions, and those who did not (only completed the tests). We can give as a main conclusion that the higher the level of involvement of the student, the better results he/she achieves.

Overall, this methodology has proved to be a promising mechanism to encourage students to participate in the teaching-learning process, and to effectively enhance their academic performance. It is for this reason that, in our faculty, we are slowly introducing this method in more and more courses each academic year.

## Appendix A. ECTS grade equivalence

To understand the Spanish grade system, we present Table A.1. This table is only for guidance, to facilitate the understanding of the Spanish grade system.

Table A.1 ECTS grade equivalence

| Percentages | Grade | Conversion | Numeric value |
|---|---|---|---|
| 95% - 100% | MH | A+ | 4 |
| 85% - 95% | SB | A | 3 |
| 65% - 85% | NT | B, C | 2 |
| 50% - 65% | AP | D | 1 |
| 0% - 50% | SS | E, F | 0 |

## Acknowledgments

The authors are very grateful to Dr. Juan Carlos Fabero, Dr. Katzalin Olcoz and Dr. Jose Manuel Velasco for participate actively in the project and providing us the data of the grades obtained by their students. We would like also to especially mention *Decanato de la Facultad de Informatica (UCM)* for helping us to carry out this study.

## References

Adeyinka T, Mutula S (2010) A proposed model for evaluating the success of WebCT course content management system. Computing Human Behavior, 26,1795–1805.

Apple (2011) Apple distinguished educators. http://ade.apple.com/application/europe/es, 2011. Last access July 2012.

Arnold S, Fisler J (2010) OLAT: The Swiss Open Source Learning Management System. In Proceedings of the 2010 International Conference on e-Education, e-Business, e-Management and e-Learning (pp. 632–636) Washington: IEEE Computer Society.

Barak M, Rafaeli S (2004) On-line question-posing and peer-assessment as means for web-based knowledge sharing in learning, International Journal of Human-Computer Studies, Volume 61, Issue 1, July 2004, Pages 84-103.

Boekaerts M, Minnaert (2003) Assessment of students´feelings of autonomy, competence, and social related-ness: a new approach to measuring the quality of the learning process through self-assessment. In M.S.R. Segers, F.J.R.C. Dochy , E.C. Cascallar, Optimizing New Methods of Assessment: In  Search of Quality and Standards. Dordecht : Kluwer Academic Publishers.

Brooks JG, Brooks MG (1993) In search of understanding: The case for the constructivist classroom . In Alexandria, V A: Association for Supervision and Curriculum Development. In Caprio, M.W

Carrick-Simpson B,  Armatas C (2003) Students' Interaction whit On-line Learning Activities: the Role of Study Strategies and Goals and Computer Attitudes.  In. Proceedings of the 20th Annual Conference of the Australasian Society for Computers in Learning in Tertiary Education (pp. 104-114) Australia: Adelaide.

Cavalli A, Maag S, Papagiannaki S, Verigakis G (2005) From UML models to automatic generated tests for the dotLRN e-learning platform. Electronic Notes in Theoretical Computer Science, 116, 133–144.

Chin C, Brown DE (2002) Student-generated questions: A meaningful aspect of learning in science. International Journal of Science Education, Vol. 24, Iss. 5, 2002.

Cohen R (1983) Self-generated questions as an aid to reading comprehension. The Reading Teacher, 36, 770–775.

Coldwell J, Craig A, Goold A(2011) Using e-Technologies for Active Learning.  Interdisciplinary Journal of Information, Knowledge and Management, Vol 6, pages 95-106.

Costa J, Caldeira H, Gallastegui J R, Otero J (2000) An analysis of question asking on scientific texts explaining natural phenomena. Journal of Research in Science Teaching, 37, 602–614.

Deci E L, Ryan RM (2010) Intrinsic Motivation. John Wisley , Sons, Inc. All.

Dreher MJ, Gambrell LB (1985) Teaching children to use a self-questioning strategy for studying expository prose. Reading Improvement, 22, 2–7.

Eow Yee L, Wan Zahbte WA, Rosnaini M, Roselan B (2010) Computer games development experience and appreciative learning approach for creative process enhancement.  Computers , Education, 55 (3), 1131-1144.

Fried CB (2008) In-class laptop use and its effects on student learning. Computers , Education, 50, 906-914.

Harasim, LM (1990) Online education: Perspectives on a new environment. New York: Praeger

HotPotatoes (2012) http://hotpot.uvic.ca/. Last access July 2012.

Joglar N, Martin D, Colmenar J M, Martínez I (2009) iTest: An Online Tool for Assessment and Self-Assessment in Mathematic.  In Proceedings of the 2009 11th IEEE International Symposium on Multimedia (pp. 663–668) Washington: IEEE Computer Society.

Keeling R (2006) Bologna Process and the Lisbon Research Agenda: the European Commission's expanding role in higher education discourse. European Journal of Education, 41(2), 203-223.

King A. (1994) Guiding knowledge construction in the classroom: effects of teaching children how to question and how to explain. American educational research journal, 30, 338–368.

Kuo LH, Tseng TH, Yang HJ, Yang HH (2010) Design a Moodle synchronous learning Activity. WTOS, 9, 409–421.

Lee RA, DePue J (2010) Using baldrige method frameworks, excellence in higher education standards, and the Sakai CLE for the self assessment process. In Proceedings of the 38th annual fall conference on SIGUCCS (pp. 165-170) New York: ACM.

Leight H, Saunders C, Calkins R, Withers M (2012) Collaborative Testing Improves Performance but Not Content Retention in a Large-Enrollment Introductory Biology Class. CBE Life Sci Educ vol. 11 no. 4, pp. 392-401

Marbach-Ad G, Sokolove PG (2000) Can Undergraduate Biology Students Learn to Ask Higher-Level Questions?. Journal of Research in Science Teaching, 37, 854-870.

Martens RL, Gulikers J, Bastiaens T (2004) The impact of intrinsic motivation on e-learning in authentic computer tasks. Journal of computer Assisted learning, 20, 368-376.

Marton F, Säljö R (1984) Approaches to learning. In F. Marton, D. Hounsell , N. Entwistle, The Experience of Learning (pp. 36-55) Edinburgh: Scottish Academic Press.

Mc Daniel MA, Fisher RP (1991) Tests and test feedback as learning sources. Contemporary Educational Psychology, 16, 192–201.

Montfort D B, Brown S, (2012) What Do We Mean by Cyberlearning: Characterizing a Socially Constructed Definition with Experts and Practitioners. Journal of Science Education and Technology. Doi:10.1007/s10956-012-9378-8.

O'Donnell AM (1999) Structuring dyadic interaction through scripted cooperation. In A. M. O'Donnell , A. King (Eds.), Cognitive perspectives on peer learning (pp. 179–196) Mahwah NJ: Lawrence Erlbaum Associates.

Oncu S, Cakir H (2011) Research in online learning environments: Priorities and methodologies. Computers , Education, 57, 1098-1108.

Pollard JK(2006) Student reflection using a Web-based Quiz. In Proceedings of the 7th IEEE International Conference on Information Technology based higher Education and Training (pp. 871-874) Australia.

Raeff C (2009) Teaching and learning through asking basic disciplinary questions: Examples from developmental psychology. Pedagogy and the Human Sciences, No. 1, pp. 28-37

Rao SP, Collins HL, DiCarlo SE (2002) Collaborative testing enhances student learning. Adv Physiol Educ, vol. 26 no. 1, pp. 37-41.

Rosenshine B, Meister C, Chapman S (1996) Teaching Students to Generate Questions: A Review of the Intervention Studies. REVIEW OF EDUCATIONAL RESEARCH, vol. 66 no. 2, pp. 181-221

Slusser SR, Erickson RJ (2006), Group Quizzes: An Extension of the Collaborative Learning Process. Teaching Sociology, vol. 34 no. 3, pp. 249-262.

Taboada A, Guthrie JT (2006) Contributions of Student Questioning and Prior Knowledge to Construction of Knowledge from Reading Information Text. Journal of Literacy Research, vol. 38 no. 1, pp. 1-35, 2006

Taplin M (2000) Problem-based learning in distance education: Practitioners' beliefs about an action learning project. Distance Education, 21(2), 284-307.

Tiernan P(2013) A study of the use of Twitter by students for lecture engagement and discussion. Journal of Education and Information Technologies, pages 1-18, Springer

Van den Broek P, Tzeng Y, Risden K, Trabasso T,Basche P. (2001) Inferential questioning: Effects on comprehension of narrative texts as a function of grade and timing. Journal of Educational Psychology, 93, 521–529.

Fu-Yun Y,(2011) Multiple peer-assessment modes to augment online student question-generation processes. Computers & Education, Volume 56, Issue 2, Pages 484-494.

Yarnall L, Shechtman N, Penuel W R (2006) Using Handheld Computers to Support Improved Classroom Assessment in Science: Results from a Field Trial. Journal of Science Education and Technology, Vol 5(2), pages 142-158

Yu-Feng L, Pin-Chuan L, Chun-Ling H (2012) An Approach to Encouraging and Evaluating Learner's Knolowledge Contribution in Web-Based Collaborative Laerning. Journal of Educational Computing Research, Vol 47(2), pages 107-135

Fong-Ling F, Ya-Ling W, His-Chuan H (2009) An investigation of coopetitive pedagogic design for knowledge creation in Web-based learning. Computers & Education, vol 53, pp. 550-562